\documentclass{article}
\usepackage[preprint]{spconf,amsmath,graphicx,url}

\begin{document}
\ninept

\onecolumn
\noindent IEEE Copyright Notice:\\

\textcopyright 2019 IEEE. Published in the IEEE 2019 International Conference on Acoustics, Speech, and Signal Processing (ICASSP 2019), scheduled for 12-17 May, 2019, in Brighton, United Kingdom. Personal use of this material is permitted. However, permission to reprint/republish this material for advertising or promotional purposes or for creating new collective works for resale or redistribution to servers or lists, or to reuse any copyrighted component of this work in other works, must be obtained from the IEEE. Contact: Manager, Copyrights and Permissions / IEEE Service Center / 445 Hoes Lane / P.O. Box 1331 / Piscataway, NJ 08855-1331, USA. Telephone: + Intl. 908-562-3966.

\twocolumn
\title{A UNIFIED NEURAL ARCHITECTURE FOR INSTRUMENTAL AUDIO TASKS}

\name{Steven Spratley, Daniel Beck, and Trevor Cohn}
\address{School of Computing and Information Systems\\
	The University of Melbourne, Victoria, Australia}

\maketitle
\setcounter{page}{1}
\copyrightnotice{\copyright\ IEEE 2019}
\toappear{To appear in {\it Proc.\ ICASSP 2019, May 12-17, Brighton, UK}}

\begin{abstract}
Within Music Information Retrieval (MIR), prominent tasks --- including pitch-tracking, source-separation, super-resolution, and synthesis --- typically call for specialised methods, despite their similarities. Conditional Generative Adversarial Networks (cGANs) have been shown to be highly versatile in learning general image-to-image translations, but have not yet been adapted across MIR. In this work, we present an end-to-end supervisable architecture to perform all aforementioned audio tasks, consisting of a WaveNet synthesiser conditioned on the output of a jointly-trained cGAN spectrogram translator. In doing so, we demonstrate the potential of such flexible techniques to unify MIR tasks, promote efficient transfer learning, and converge research to the improvement of powerful, general methods. Finally, to the best of our knowledge, we present the first application of GANs to guided instrument synthesis.
\end{abstract}

\begin{keywords}
music information retrieval, generative adversarial network, audio modelling, synthesis
\end{keywords}

\section{Introduction}
\label{sec:intro}

Within Music Information Retrieval (MIR), the prominent tasks of \textbf{pitch-tracking}, \textbf{source-separation}, \textbf{super-resolution}, and \textbf{synthesis} are usually treated as distinctly different tasks, employing widely varying techniques \cite{RN146,ward2018sisec}. We suspect some redundancy in this approach; humans can listen for particular instruments in an ensemble, track their behaviour, and imagine `hearing' them without external stimuli, all as a by-product of learning each instrument's sound. Consequently, we expect that methods for processing audio should be adept in several such tasks, jointly in a multi-task learning setting.

Despite requiring different methods, time-frequency spectrograms are central to most current techniques. Recently, \textit{WaveNets} \cite{RN176} were shown to reconstruct such spectrograms more accurately than the long-standing Griffin-Lim algorithm \cite{RN178}, yet this has barely been applied outside of speech processing. Meanwhile, image translation models such as \textit{Generative Adversarial Networks} (GANs) have achieved wide success; several papers have investigated their application to speech audio \cite{RN185,Gao2018VoiceIU}. Yet, little work has been done to investigate their general application across multiple tasks, or to instrumental audio. Therefore, to frame the aforementioned tasks as the same underlying problem is an attractive pursuit, potentially allowing research to leverage powerful new methods, efficiently train multi-task models, and converge on improving the one technique.

\begin{figure}
	\begin{minipage}[b]{1.0\linewidth}
		\centering
		\centerline{\includegraphics[width=8.5cm]{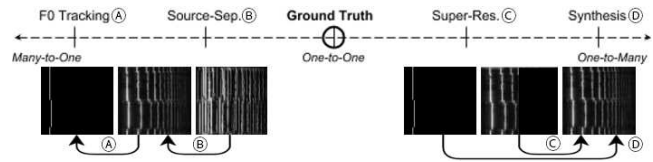}}
	\end{minipage}
	\caption{Framing prominent instrumental audio tasks as translation. Illustrated with spectrograms; frequencies can be distilled in operations to the left, or added, to the right.}
	\label{translation}
	\vspace{-0.3cm}
\end{figure}

Music is sometimes conceived of as language without semantics; one cannot converse with music about \textit{concepts}, yet it displays syntax and organisation at the levels of rhythm and harmony. Therefore, one ought to be able to consider its translation. As a sentence may be embellished or summarised, the same notes may be performed by an ensemble or a single instrument, differentiated by the number of frequencies superposed in the mix. Translation in this limited sense becomes the act of adding or stripping away frequencies (Figure \ref{translation}). On one extreme, pitch-tracking can be thought of as distilling the fundamental frequency, $F_0$. Its opposite, synthesis, is therefore taking $F_0$ and introducing the rest of the spectrum. Source-separation of polyphonic audio becomes detecting the frequencies made by a single instrument and ignoring the rest, whereas super-resolution restores upper frequencies to low-fidelity recordings.

Given the proven utility of the time-frequency spectrogram in a wide range of audio tasks, the framing of these tasks as a translation problem, the emergence of image translation methods, and improved spectrogram reconstruction techniques, this paper seeks to develop a single model framework capable of each task, and their combination. Our pipeline consists of a conditional GAN that learns translations between mel-scaled spectrograms, which in turn, conditions a WaveNet synthesiser to reconstruct the final audio. This contributes a generalisable and end-to-end supervisable architecture that displays both competitive performance across a wide application, and further gains from joint multi-task learning.\footnote{Examples at \url{https://svenshade.github.io/GAN-WN/}.}
\vspace{-0.2cm}

\section{MODELS}
\label{sec:models}

\subsection{Generative Adversarial Networks and pix2pix}
\label{ssec:gans}

\begin{figure}
	
	\begin{minipage}[b]{1.0\linewidth}
		\centering
		\centerline{\includegraphics[width=8.5cm]{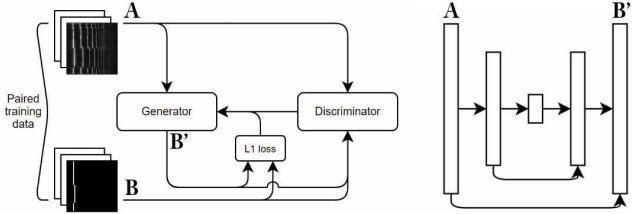}}
	\end{minipage}
	\caption{Illustration of cGAN (left) and its U-Net generator (right).}
	\label{cGAN}
\end{figure}

Generative adversarial networks (GANs) \cite{RN211} are frameworks consisting of generator and discriminator ($G$ and $D$) subnetworks that learn via competition. $D(y; \theta_{d})$ seeks to maximise the probability of correctly discerning whether $y$ was taken from real data $x{\sim}p_{data}(x)$ or generated, counterfeit data. Simultaneously, $G(z; \theta_g)$ learns to map input noise $z{\sim}p_{z}(z)$ to $x$ in order to produce increasingly convincing imitations. $G$ is never shown the data it aims to generate, instead it relies solely on $D$'s output as part of its own loss. The parameters $\theta_g$ are trained to minimise $\log{(1-D(G(z))}$, while parameters $\theta_d$ learn to maximise $\log{D(x)} + \log{(1-D(G(z))}$. Due to the codependence of the networks, the training scheme involves alternating their update phases. \textit{pix2pix} (cGAN) \cite{RN174} extends the framework by conditioning both subnetworks on images; $G$ learns to map from noise to translation, conditioned on an observation, while $D$ compares the same observation with a translation of unknown origin (ground truth or imitation). This model also incorporates L1 distance to stabilise training via objective feedback.
In pix2pix, $D$ is a convolutional classifier which assumes independence between pixels of separate patches, allowing for fewer parameters and handling images of different sizes. $G$ resembles a “U-Net” \cite{RN214} autoencoder with skip connections between mirrored layers that circumvent the information bottleneck at the centre of the network by opening up other means for decoder layers to access low-level information (e.g. edges) that might be near identical with their paired encoder layers (see Figure \ref{cGAN}). This feature is a strength of pix2pix, as there are many image translation tasks, including our own, that require some elements of the original composition to remain unchanged.

This work further demonstrates that loss functions needn't be explicitly specified by the researcher between tasks, as $D$ effectively learns a bespoke loss. It therefore presents a flexible method that can be applied to a number of image translation problems, which our research aims to exploit in the spectrogram space.

\subsection{WaveNet and Tacotron 2}
\label{ssec:wn}

A key hurdle presents itself when modelling raw audio. Capturing high-level image features (e.g. a cat) might require deeper neurons to be receptive to say, 200 pixels squared. A high-level audio feature (e.g. a bowed string) may be sustained over a second, increasing the receptive field requirement to tens of thousands of timesteps, even for a medium-fidelity signal. Although this is possible with large kernel sizes and strides, it doesn't present a way to overcome the second hurdle, which is to respect the sequential nature of audio. 
\textit{WaveNets} \cite{RN173} make use of causal, dilated convolutional layers. By both forcing connections to feed forward into later timesteps, and progressively dilating hidden layers, the resultant network can be visualised as a binary tree that slides along the data, enabling the receptive field to increase exponentially with layer depth. In this way, entire waveforms can be generated autoregressively; by classifying one timestep at a time given the existing series.

In a follow-up paper \cite{RN176}, \textit{Tacotron 2} is described as a fully neural architecture for text-to-speech, making use of WaveNet as a synthesiser conditioned on features generated by the original Tacotron network, replacing the Griffin-Lim spectrogram-to-audio algorithm. Not only does this make the system supervisable end-to-end, it also allows compressed representations such as mel-spectrograms to be reconstructed directly, reducing network complexity. 

\section{THE GAN-WN ARCHITECTURE FOR VARIOUS INSTRUMENTAL AUDIO TASKS}
\label{sec:gan-wn}

\subsection{Data and Representation}
\label{ssec:representation}

We chose to exclusively model the solo violin for three reasons. One, forming multiple `deep' models for each task is infeasible with our compute resources. Two, the violin is notoriously hard to model due to its expressive range \cite{RN169}, serving as an effective proof-of-concept for many other instruments. Three, we expect it to facilitate efficient joint modelling of tasks. For pitch-tracking and synthesis tasks, we created paired multi-sine-wave (fundamental + first 5 harmonics) and violin audio tracks using chromatic scales, software instruments, and data from the \textit{Bach10} dataset \cite{RN204}. For super-resolution, we downsampled over 12 hours of live violin recordings \cite{RN235,RN237,RN234,RN236} in order to generate the paired low-fidelity track. Finally, for source-separation we used multi-track recordings from the \textit{Bach10}, \textit{Freischutz}, and \textit{Phenicx-anechoic} datasets \cite{RN204,RN206,RN205}, and created our own synthetic multi-track data using MIDI files of Bach's \textit{Four Orchestral Suites}\footnote{\url{http://www.jsbach.net/midi/}}, played through software instruments. We further tailored MIDI files by reducing the complexity of polyphonic phrases in order to clarify connection to $F_0$, and keeping instrument sources to under five.
We chose to model at a literature-standard sample rate of 16 kHz, as the spectrum becomes increasingly sparse in upper frequencies captured by higher rates, meaning computationally-expensive diminishing returns. Once we had finalised our audio tracks, they were compressed and normalised, segmented into chunks, and processed via the short-time Fourier transform (STFT) \cite{RN178}. STFT parameters (hop size of 49 samples, and window size of 1024 and length of 640 samples) were co-determined with chunk duration (1550ms), to settle on a representation that displayed clear frequency lines when viewed as an image at the target resolution of 256x256 pixels. Finally, we mel-scaled all data to efficiently model for human frequency perception, and reclaim memory for larger kernels and layers.

\subsection{Translation}
\label{ssec:translation}

Our method extends the pix2pix method presented in \cite{RN174} (using the official public implementation) in order to fit a translation model to each of our datasets of paired spectrograms. In each case, once the generator, $G$ is properly trained, testing becomes a matter of converting audio of arbitrary length to its mel-spectrogram representation and applying $G$ convolutionally. In this domain, pitch-tracking can be thought of as semantic segmentation in the same way that a satellite image might be translated into a road map. Source-separation becomes a denoising problem, relaying patterns related to the signal and ignoring others. Super-resolution is analogous to inpainting, where the first half of the spectrogram --- the low frequencies --- is used to fill in the blank half. Finally, synthesis becomes a style transfer problem, where the model has extracted the overall harmonic `style' of an instrument, and seeks to apply this to sine-waves serving as a harmonic \textit{blueprint}. After a preliminary grid search over hyperparameters, we adopted a kernel size of 8x8 for both $G$ and $D$ networks and optimised for least-squares loss.

\subsection{Reconstruction}
\label{ssec:reconstruction}
\begin{figure}
	\begin{minipage}[b]{1.0\linewidth}
		\centering
		\centerline{\includegraphics[width=8.5cm]{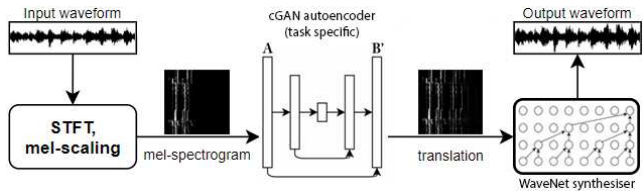}}
	\end{minipage}
	\caption{The GAN-WN architecture.}
	\label{GAN-WN}
\vspace{-0.2cm}
\end{figure}

We present three spectrogram reconstruction methods. The first involves naively rescaling frequencies before processing with Griffin-Lim, which displays noticeable artefacts and a lack of detail in higher frequencies. The second method improves on this by inserting a secondary cGAN process after rescaling, trained on ground-truth and mel-scaled-rescaled data, to further restore the linear spectrograms. The third method makes use of the WaveNet architecture to circumvent both lossy rescaling and Griffin-Lim reconstruction, locally conditioned on spectrograms to guide generation. We trained one WaveNet model on the task that had the most available data (super-resolution). By partitioning into equal test and train ($\sim$6 hours of audio each), our cGAN produced 6 hours-worth of spectrograms from unseen data, which were then paired with their ground truth audio in order to train WaveNet. In doing so, we train GAN-WN as a cascade architecture, making the reconstruction stage more robust to translation errors. We illustrate the GAN-WN architecture in Figure \ref{GAN-WN}.
Memory resources for WaveNet limited training instances to $\sim$30k steps, or 1.88s in duration, which perfectly suited the 1.55s chunk size represented by our datasets. Noticing that our first WaveNet over-emphasised harmonics, we implemented a technique to further guide audio called \textit{teacher-weighted generation}, which makes use of the original, low-fidelity signal being restored, by first interpolating it to the target sample rate, and taking a weighted average between each timestep generated by the network and the corresponding interpolated timestep, by a set factor. That prediction, $f^{-1}((f-1)x_i + y_i)$, where $x_i$ is the timestep from the interpolated signal, $y_i$ is WaveNet's current prediction, and $f$ is our weight factor, is fed back into the network to discourage errors from being propagated. We found that a factor of 20 was sufficient to generate audio that was less prone to squealing, at the cost of some detail.

\section{EXPERIMENTS AND RESULTS}
\label{sec:experiments}

After fitting a dedicated cGAN model to each of the four tasks' datasets, we evaluate on the bases of spectrogram distance as well as human audition. For measuring the former, we report both percentage difference (normalised L1 distance) as well as structural similarity (SSIM), which aims to independently consider structural information from illumination \cite{RN219}. For the latter, we hosted an online, APE-style audition test \cite{RN223} for twenty students and colleagues from the authors' personal networks. This evaluation was designed to facilitate comparison between multiple, time-aligned audio clips, by presenting the participant with a series of 20 horizontal scales, featuring sets of 4-6 audio clips corresponding to the same 10-second test instance as processed by the methods. Clip positions and labels were randomised at test time to avoid bias. Participants were tasked with auditioning each set and assessing each clip against a continuous Likert scale, across \textit{bad}, \textit{poor}, \textit{fair}, \textit{good}, and \textit{excellent} markers. This format ensured that for each instance, all of the methods were compared with each other, and with finer resolution. The test was built using the \textit{Web Audio Evaluation Tool} (WAET) \cite{RN224}.

\subsection{Harmonic Distillation: Pitch-Tracking and Source-Separation}
\label{ssec:distillation}

\begin{figure}
	\begin{minipage}[b]{1.0\linewidth}
		\centering
		\centerline{\includegraphics[width=8.5cm]{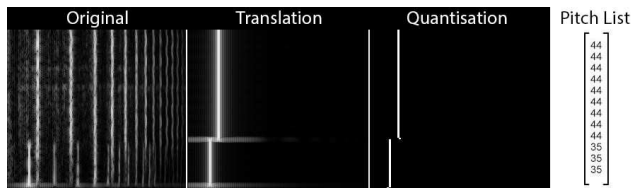}}
	\end{minipage}
	\caption{The pitch quantisation process, proceeding left to right.}
	\label{pitch}
\end{figure}

The ubiquitous baseline for pitch-tracking, \textit{pYIN} \cite{RN148}, performed poorly over our data due to several octave errors. This prompted us to use the more sophisticated audio-to-MIDI function in the commercial digital audio workstation, \textit{Ableton Live 9}. Unlike linearly-scaled spectrograms, training the model on the mel-scale ensures that the columns (filterbanks) closely correlate with musical pitch; understanding cGAN's pitch predictions was a simple matter of recording the filterbank number of the brightest pixel in each row (Figure \ref{pitch}). Over our test set, we tallied correctly predicted notes from both our method and the baseline, but to credit notes that were mistimed yet otherwise pitched properly, and investigate alignment of note onsets and offsets, we also tallied incorrectly classified timesteps (pixel rows), which we report as mean error per spectrogram (Table \ref{pitchres}). Each timestep was considered incorrect if it was neither in, nor neighbouring, the correct row.

\begin{table}
	\begin{center}
		\begin{tabular}{c|c|c|c|c}
			\textbf{Method} & \textbf{Correct} & \textbf{Total} & \textbf{Precision} & \textbf{Mean error}\\
			\hline
			Ableton & 109 / 130 & 157 & 69.43\% & 32.69 \\
			cGAN & \textbf{119 / 130} & \textbf{122} & \textbf{97.54\%} & \textbf{17.35}
			\end{tabular}
			\caption{Pitch-tracking results, including correct/total notes found.}
			\label{pitchres}
		\end{center}			
\vspace{-0.7cm}

\end{table}

For source-separation, we used both blind and supervised baselines; the non-negative matrix factorisation (NMF) method implemented in the \textit{Flexible Audio Source Separation Toolbox} \cite{RN226}, and the convolutional neural network-based method (CNN) detailed in \cite{RN151}. The latter technique is chosen due to the high similarity of training data it shares with our method (Bach10 and pre-recorded violin samples), providing a more challenging baseline. The test piece --- a Bach piano and violin duet --- was chosen to better test how the methods fare on real-world data; neither the piano, music, nor number of sources, is featured in the training data for either method. See Table \ref{spectres} for spectrogram results. For the listening tests in Figure \ref{listres}, the prompt: ``Overall Success: Removing Piano, Keeping Violin'' was given to participants to rate against.

\begin{table}
	\begin{tabular}{lc|c|c|ccc}
		& \multicolumn{3}{c|}{\textit{Source-Separation}} 
		& \multicolumn{3}{c}{\textit{Super-Resolution}} \\
		& \textbf{cGAN} & \textbf{NMF} & \textbf{CNN} & \multicolumn{1}{c|}{\textbf{cGAN}} & \multicolumn{1}{c|}{\textbf{Cubic}} & \textbf{Linear} \\
		\hline
		\multicolumn{1}{c|}{\% error} & \textbf{4.21} & 4.96 & 5.24 & \multicolumn{1}{c|}{\textbf{1.65}} & \multicolumn{1}{c|}{2.42} & 2.63 \\
		\multicolumn{1}{c|}{SSIM}     & \textbf{0.52} & 0.50 & 0.47 & \multicolumn{1}{c|}{\textbf{0.68}} & \multicolumn{1}{c|}{0.53} & 0.51           
	\end{tabular}
	\caption{Spectrogram performance over source-separation and super-resolution test sets.}
	\label{spectres}
\end{table}
\vspace{-0.1cm}
\subsection{Harmonic Addition: Super-Resolution and Synthesis}
\label{ssec:addition}

\begin{figure}
	\begin{minipage}[b]{1.0\linewidth}
		\centering
		\centerline{\includegraphics[width=8.5cm]{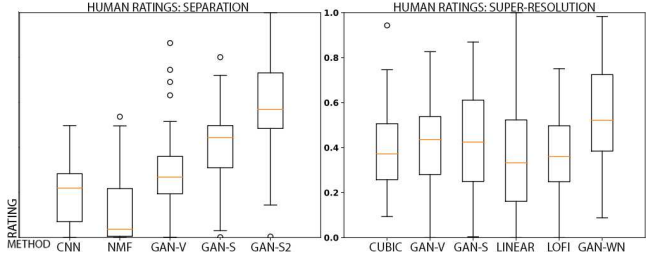}}
		\vspace{-0.3cm}
	\end{minipage}
	\caption{Listening test plots. GAN-V refers to naive reconstruction, while GAN-S \& S2 use 1 \& 2 secondary cGAN stages respectively.}
	\label{listres}
\vspace{-0.2cm}
\end{figure}

\begin{figure}
	\begin{minipage}[b]{1.0\linewidth}
		\centering
		\centerline{\includegraphics[width=8.5cm]{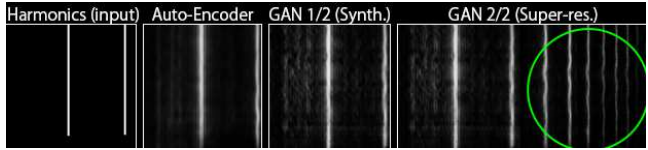}}
		\vspace{-0.3cm}
	\end{minipage}
	\caption{Comparing synthesis models. Note the realistic trailing vibrato waves reproduced (encircled).}
	\label{synthesis}
\vspace{-0.4cm}
\end{figure}

For super-resolution, as in \cite{DBLP:journals/corr/abs-1708-00853} we used interpolative baselines (linear, cubic b-spline) \cite{RN163}, which were compared against our pipeline using all three reconstruction methods (Table \ref{spectres}). While \cite{DBLP:journals/corr/abs-1708-00853} itself provides a recent, informed technique, interpolation is sufficient for proof-of-concept, keeping to our aim of investigating the generality of our method, and mitigating the cost of training another baseline. We tested over solo violin audio from a different recording environment, downsampled to 4 kHz in order to perform 4x upsampling back to 16 kHz. We include the outcomes of the two super-resolution listening tests in Figure \ref{listres}; participants were given the prompt ``Overall Quality''. While pixel-wise loss is a useful metric for distillation tasks, it becomes decreasingly useful the more we introduce new frequencies, where the real aim is to fill in sound that is believable, instead of meeting ground truth. We chose to evaluate synthesis by ablation; we remove $D$ and train a baseline autoencoder on L1 distance alone. We train both methods over the synthesis dataset, this time making use of the aforementioned multi-sine-wave (harmonics) track; the single sine-wave track stalled training due to its sparsity. We also use our synthesis model in conjunction with the super-resolution model to avoid redundancy; we train the former on the simpler task of translating harmonics to 4 kHz violin audio, and then leverage the latter, trained on far more data, to inpaint the rest (Figure \ref{synthesis}). During training, the capabilities of each model became clear, with the autoencoder requiring three times as many epochs as the cGAN model before reproducing clear noise + harmonics patterns.

\subsection{Jointly Learning to Track, Separate, and Enhance}
\label{ssec:joint}

\begin{figure}
	\begin{minipage}[b]{1.0\linewidth}
		\centering
		\centerline{\includegraphics[width=8.5cm]{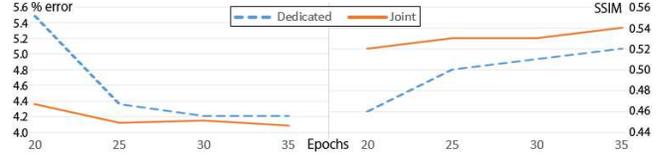}}
		\vspace{-0.3cm}
	\end{minipage}
	\caption{Comparing joint vs. dedicated models.}
	\label{joint}
\vspace{-0.2cm}
\end{figure}

As stated earlier, there ought to be potential for significant transferable learning, with the acquisition and refinement of an instrument's audio profile facilitating multiple tasks. To pursue this, we formed a combined dataset of equal parts of our original sets, and trained a joint model by increasing the input channels of our cGAN to three (we collapsed synthesis and super-resolution to the one task, \textit{enhancement}). We chose to compare this model against the dedicated source-separation model, given its complexity. We believe our results encourage such an approach (Figure \ref{joint}), confirming our expectation that kernels not only ought to be shared efficiently between tasks, but also cross-benefit, as shown by the increased performance our joint model achieved, in fewer iterations, despite only processing the same amount of task-specific data as the dedicated model.

\subsection{Discussion}
\label{ssec:discussion}

We observe promising results; the models outperformed baselines, and subjective tests displayed clear preference for our method. For pitch-tracking, a common error made by \textit{Ableton} was incorrectly timed note onsets (difficult for continuous-excitation instruments); in its defence, our method was trained to handle solo violin audio, and the metric we used only registered one correct note per ground truth note, which penalised Ableton's subdivided notes. Across distillation tasks, cGAN showed greater precision in the frequencies it detected, sometimes trading off detail in the high frequencies. Further tuning of our loss function could help approach CNN's accuracy in that regard, although CNN underperformed overall, conceivably due to it being trained to separate a set four sources. We also tested Signal-to-Distortion and Source-to-Interference ratios (SDR, SIR) for source-separation, in order to better quantify performance. However, we noticed a severe drop in SIR (from 25dB, down to 16dB) after our second cGAN restoration, as well as non-competitive SDR scores. Curiously, this is not consistent with the results of the subjective listening test, and we hypothesise that such success is due to the generative nature of the GAN process; it is liable to fill in statistically-consistent detail, even if such detail is not present in the signal. This marked difference in the way our technique generates audio compared to baselines makes it tricky for objective comparison, and we encourage future research in perceptual metrics for GAN-based MIR systems. This finding reveals a weakness in our technique as compared to specialised approaches, as source-separation arguably has a ground-truth that needs to be uncovered. To re-state; our approach is not to overtake state-of-the-art, but to explore and give credence to the efficient research strategy of investing in general techniques that model with less redundancy in training, since any data might increase performance in all tasks.

\begin{figure}
	\begin{minipage}[b]{1.0\linewidth}
		\centering
		\centerline{\includegraphics[width=8.5cm]{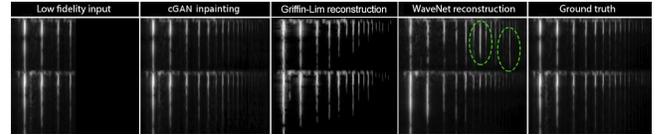}}
		\vspace{-0.3cm}
	\end{minipage}
	\caption{Examples from our reconstructive processes on super-resolution. Whistling artifacts (encircled) introduced by WaveNet, and note the thresholding degradation from descaling \& Griffin-Lim.}
	\label{reconstruction}
	\vspace{-0.33cm}
\end{figure}

For addition tasks, sparsity issues (mapping too few frequencies to complex spectra) were aided by mel-scaling. WaveNet was able to model noise components better than Griffin-Lim, yet was let down by artefacts; we hypothesise this is analogous to positive feedback in language models, where a particular phrase predictably sparks others, propagating error. This was alleviated by teacher-weighted generation, at the cost of some detail in the upper spectrum. See Figure \ref{reconstruction} for a comparison of spectrograms output by the different processes. Regarding algorithm performance, cGAN translation was more than an order of magnitude faster than the reconstruction stage, at $\sim$19x realtime. Our implementation of WaveNet\footnote{Based largely on \url{github.com/r9y9/wavenet_vocoder}.} was slower, only able to produce $\sim$50 raw audio samples a second. Griffin-Lim reconstruction was much faster ($\sim$0.78x realtime), allowing $\sim$0.5x realtime speed to complete any task using the cGAN-S method overall, and a $\sim$4x speedup compared to source-separation using our NMF baseline. All models were trained using an NVIDIA GTX 1080ti GPU.

\vspace{-0.15cm}
\section{CONCLUSIONS AND FUTURE DIRECTIONS}
\label{sec:conclusions}

In this work, we have demonstrated how recent innovations in speech and image processing can impact diverse MIR problems, contributing to their unification in theory and practice. Our approach, GAN-WN, is a general, supervisable method that can be jointly-trained, retaining competitive performance across most tasks. By making use of STFT phase data, lossy spectrogram transformations might become redundant in future iterations, reducing artefacts. Finally, adding expressive control of synthesis, and pursuing multi-instrument modelling, are both promising extensions.

\vfill
\pagebreak

\bibliographystyle{IEEEbib}
\bibliography{PaperBib}

\end{document}